\documentclass{acm_proc_article-sp}
\usepackage{graphicx}
\usepackage{wrapfig}

\long\def\comment#1{}
\newcommand{\commentout}[1]{}

\comment{
\setlength{\topmargin}{0.0in}

\setlength{\headheight}{0.0in}

\setlength{\headsep}{0.0in}

\setlength{\textheight}{9in}

\setlength{\oddsidemargin}{0in}

\setlength{\textwidth}{6.5in}

\setlength{\parindent}{0in}

\setlength{\parskip}{2ex}

\renewcommand{\baselinestretch}{1}

}
\newcommand{\denselist}{
      \setlength{\itemsep}{0pt}
      \setlength{\parsep}{1.5pt}
      \setlength{\topsep}{1.5pt}
      \setlength{\parskip}{2pt}
      \setlength{\partopsep}{0pt}
      \setlength{\labelwidth}{1em}
      \setlength{\labelsep}{0.5em} }

\newcommand{\bdesc}{\begin{description}\denselist}
\newcommand{\edesc}{\end{description}}

\newcommand{\figref}[1]{Figure~\ref{#1}}
\newcommand{\tabref}[1]{Table~\ref{#1}}

\begin{document}

\comment{\title{\LARGE \bf Social Networks and Social Information
Filtering on Digg}
\author{Kristina Lerman\\
University of Southern California \\
Information Sciences Institute\\
4676 Admiralty Way\\
Marina del Rey, California 90292\\
lerman@isi.edu }

}

\title{Social Networks and Social Information Filtering on Digg
}
\numberofauthors{1}

\author{
\alignauthor
Kristina Lerman\\
       \affaddr{University of Southern California }\\
       \affaddr{Information Sciences Institute} \\
       \affaddr{4676 Admiralty Way}\\
       \affaddr{Marina del Rey, California 90292}\\
       \email{lerman@isi.edu}
}

\maketitle 

\begin{abstract}
The new social media sites --- blogs, wikis, Flickr and Digg, among
others --- underscore the transformation of the Web to a
participatory medium in which users are actively creating,
evaluating and distributing information. Digg is a social news
aggregator which allows users to submit links to, vote on and
discuss news stories. Each day Digg selects a handful of stories to
feature on its front page. Rather than rely on the opinion of a few
editors, Digg aggregates opinions of thousands of its users to
decide which stories to promote to the front page.

Digg users can designate other users as ``friends'' and easily track
friends' activities: what new stories they submitted, commented on
or read. The friends interface acts as a \emph{social filtering}
system, recommending to user stories his or her friends liked or
found interesting. By tracking the votes received by newly submitted
stories over time, we showed that social filtering is an effective
information filtering approach. Specifically, we showed that (a)
users tend to like stories submitted by friends and (b) users tend
to like stories their friends read and liked. As a byproduct of
social filtering, social networks also play a role in promoting
stories to Digg's front page, potentially leading to ``tyranny of
the minority'' situation where a disproportionate number of front
page stories comes from the same small group of interconnected
users. Despite this, social filtering is a promising new technology
that can be used to personalize and tailor information to individual
users: for example, through personal front pages.

\end{abstract}


\comment{\noindent \textbf{Keywords:}Social Network analysis;
collaborative filtering; social filtering }

\section{Introduction} The label ``social media'' has been
attached to many Web sites --- blogs, MySpace, Flickr, del.icio.us,
Wikipedia --- whose content is primarily user driven. The recent
rise of social media sites underscores the transformation of the Web
and how it is being used. Rather than searching for and passively
consuming information found on Web pages, users are now actively
creating, evaluating and distributing information. Newer scripting
technologies and software tools allow anyone to seamlessly add
content to a Web site --- a new blog entry, or a change to an
existing article, an image, a link, a vote or feedback comment ---
without being familiar with HTML or the underlying technologies used
by that site. Most of the sites also include a social networking
component, which enables users to build personal social networks by
designating other users as ``friends'' or ``contacts'' in order to
gain access to friends' activities. For example,
Flickr~\cite{flickrurl} allows users to see in real time new images
posted by friends. Another distinctive feature of the social media
sites is their transparency. Every username, every descriptive tag
is a hyperlink that can be used to navigate the site, and unless it
has been designated private, all content is publicly viewable and in
some cases, modifiable.

Many Web sites that provide information (or sell products or
services) use collaborative filtering technology to suggest relevant
documents (or products and services) to its users. Amazon and
Netflix, for example, use collaborative filtering to recommend new
books or movies to its users. Collaborative filtering-based
recommendation systems~\cite{Konstan97grouplens} try to find users
with similar interests by comparing their opinions about products.
They will then suggest new products that were liked by other users
with similar opinions. Recommender systems based on \emph{social
filtering}, on the other hand, suggest new products or documents
simply based on whether the user's designated friends found these
products or documents interesting. Researchers in the past have
recognized that social networks present in the user base of the
recommender system can be induced from the explicit and implicit
declarations of user interest, and that these social networks can in
turn be used to make new recommendations~\cite{perugini04}. To the
best of our knowledge, social media sites are the first systems to
directly use social networks for social filtering.

In this paper we show that social filtering on Digg, a social news
aggregator, is an effective recommendation system. Specifically, we
show that Digg users tend to be interested in the news stories their
friends find interesting. We also study the effect social filtering
has on the organization of stories on Digg, including unintended
consequences such as ``tyranny of the minority.'' We compare Digg
with Reddit, another social new aggregator that, unlike Digg, uses
collaborative filtering to recommend news stories to its readers.
Reddit's type of filtering appears to be much weaker, promoting
stories that users do not find interesting. Although social
filtering, as practiced by Digg, has recently come under fire, we
believe it to be a promising technology that will lead to new
generation of personalization and recommendation algorithms.

\begin{figure*}[tbhp]
  \center{\includegraphics[width=6in]{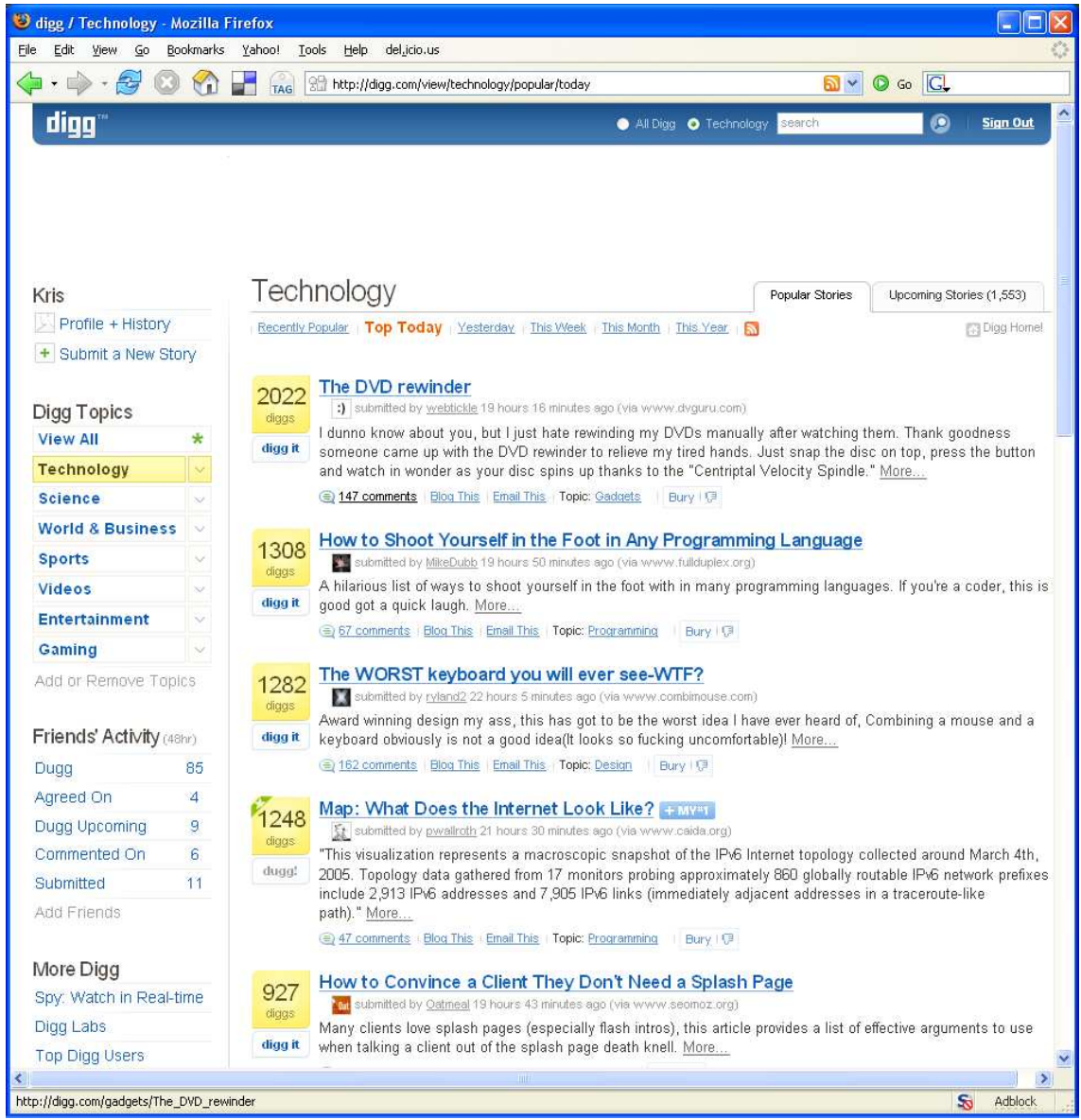}\\}
  \caption{Digg.com homepage showing front page technology stories}\label{fig:homepage}
\end{figure*}

\section{Structure of Digg}

Digg~\cite{diggurl} is arguably one of the most successful social
news aggregators. Its functionality is very simple. Users submit
links to stories they find online, and other users vote on these
stories.
When a story gets enough positive votes, or diggs, it is promoted to
the front page. The front page is what users see on the Digg home
page, while the newly submitted stories are less visible, being
``hidden'' in the Upcoming stories pages.

A typical Digg page is shown in \figref{fig:homepage}. Each contains
a list of 15 stories. The stories are in reverse chronological order
of being submitted (for the upcoming stories queue) or promoted (for
the front page stories), with most recent stories appearing at the
top. The story's title is a link to the source, while clicking on
the number of diggs takes one to the page describing the story's
activity on digg: the discussion around it, the list of people who
dugg it, etc. Digg also allows users to designate other users as
friends. Digg makes it easy to track friends' activities. The left
column on the home page summarizes the number of stories the friends
have submitted, commented on or liked recently. It even has a handy
feature to see the stories at least two friends have liked (``agreed
on''). All these stories are also are flagged with a green ribbon
(see fourth story in \figref{fig:homepage}) making them easy to
spot. Tracking activities of friends is common feature in many
social media sites and is one of the major draws attracting users to
these sites. It offers a new paradigm for interacting with
information --- social filtering. Rather than actively searching for
new interesting content, or subscribing to a set of predefined
topics, users can now put other people to task of finding and
filtering information for them.


Digg selects a handful of stories each day to feature on its front
page. Getting to the front page is important to users, because it
increases the story's visibility (most people who go to Digg only
read the front page stories), as well as the visibility of the user
who submitted the story. In fact, Digg ranks users based on how many
of their stories made it to the front page, and improving one's rank
has become a competitive sport. Although the exact formula for how a
story is promoted to the front page is kept secret, so as to prevent
users from ``gaming the system'' to promote bogus stories, it
appears to take into account the number of diggs a story gets and
the rate at which it gets them. The mechanism by which the stories
are promoted, therefore, does not depend on the decision of one or
few editors, but emerges from the activities of many users. We are
interested in studying the mechanism by which such consensus emerges
and the role social networks play in them.


\section{Dynamics of diggs} \label{sec:dynamics}

In order to see how consensus emerges from independent decisions
made by many users, we tracked both new and front page stories in
the technology category. We collected data by scraping Digg site
with the help of Web wrappers, created with tools provided by Fetch
Technologies:

\begin{description}
  \item[digg-frontpage] wrapper extracts a list of stories
from the first 14 pages of the home page. For each story, it
extracts submitter's name, story title, time submitted, number of
diggs and comments the story received.

  \item[digg-all] wrapper extracts a list of stories
from the first 20 pages in the Upcoming stories queue. For each
story, it extracts the submitter's name, story title, time
submitted, number of diggs and comments the story received.

  \item[digg-with-history] wrapper extracts the same information as
  digg-frontpage wrapper, along with the list of the first 216 users
  who dugg the story.

  \item[top-users] wrapper extracts information about the first 1020 recently active users.
  Since Digg ranks users by how many stories they have on the front page,
  we collect information about 1020 of the top ranked users.
  For each user, it extracts the number of stories
  that user has submitted, commented on, and dugg; number of stories that have been promoted to
  the front page; number of profile views; time account was
  established; users's rank; the list of
  friends (contacts), as well as reverse friends or ``people who
  have befriended this user.''
\end{description}

\emph{Digg-frontpage} and \emph{digg-all} wrappers were executed
hourly over a period of a week in May 2006. \emph{Top-users} wrapper
was executed at the same time to gather a snapshot of the social
network of the top Digg users.


\begin{figure*}[tbh]
\begin{tabular}{cc}
  \includegraphics[height=1.9in]{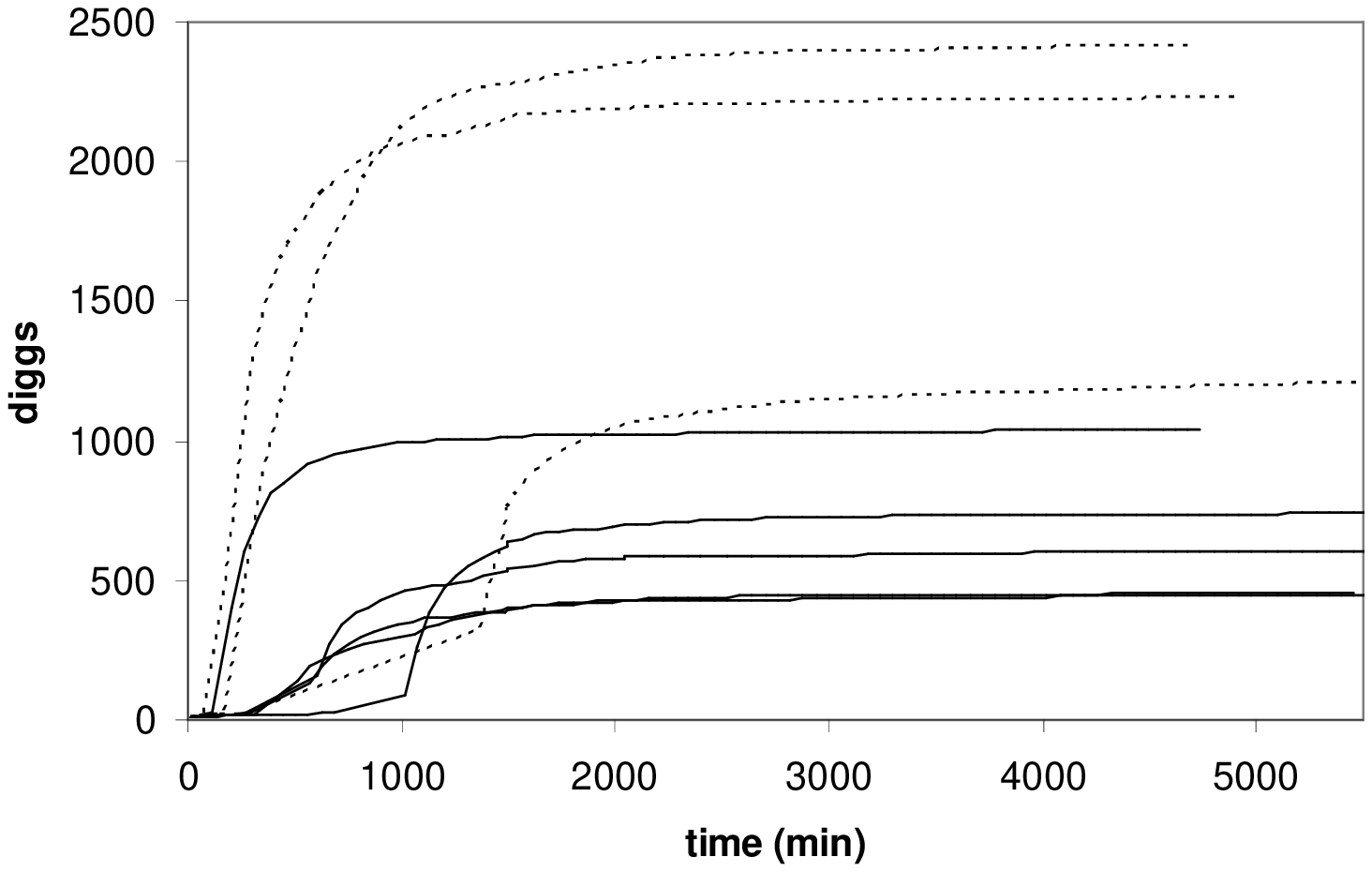}  &
  \includegraphics[height=1.9in]{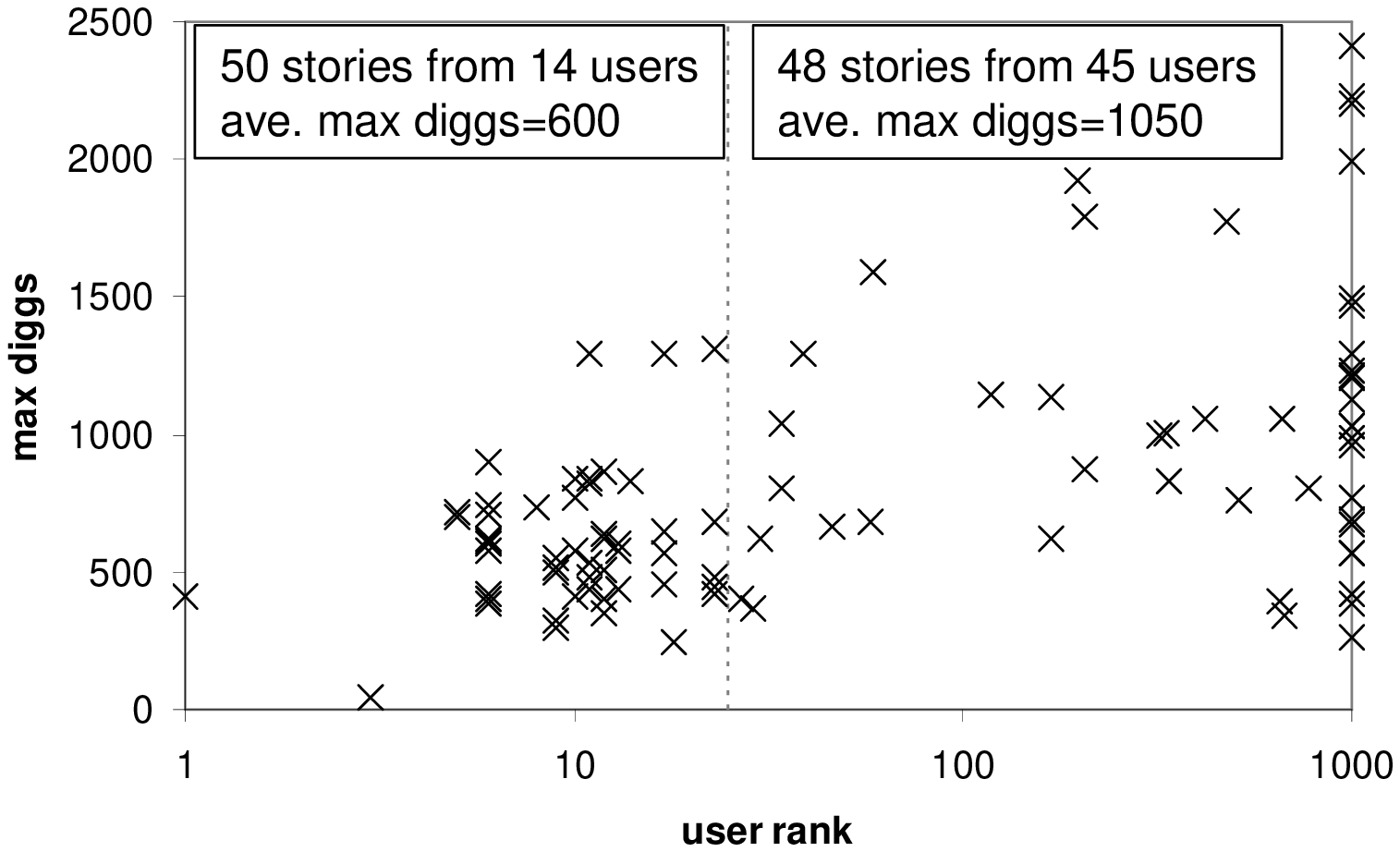}  \\
  (a) & (b) \\
\end{tabular}
\caption{(a) Dynamics of ratings (diggs) of select stories that have
been promoted to the Digg front page. Dashed lines correspond to
stories submitted by users whose rank was greater than 1020, while
solid lines correspond to stories submitted by users whose rank was
less than 35. (b) Maximum number of diggs attained by a story during
the period of observation vs submitter's rank. Symbols on the right
axis correspond to low-rated users with rank$>1020$.}
\label{fig:diggs}
\end{figure*}

We identified stories that were submitted to Digg over the course of
approximately one day and followed these stories over a period of
six days. Of the $2858$ stories that were submitted by $1570$ users
during this time period, only 98 stories by 60 different users made
it to the front page. \figref{fig:diggs}(a) shows evolution of the
ratings (number of diggs) of select stories. The basic dynamics of
all the stories appears the same. A story accrues diggs at some
rate. Once it is promoted to the front page, it accumulates diggs at
a much faster rate. \comment{The growth in the number of diggs is
impressive, considering that a front page story gets 300 views for
every digg it receives \cite{blog article}.} As the story ages,
accumulation of new diggs slows down, and the story's rating
saturates at some value. We will call the maximum diggs a story
accrues its ``interestingness'', as it reflects how interesting the
story is to the general audience.

It is worth noting that top rated users are not submitting stories
that get the most diggs. This is shown graphically in
\figref{fig:diggs}(a) where stories submitted by low-rated users
(with rank$>1020$) are shown as dashed lines, while solid lines
represent stories submitted by top-rated users.
\figref{fig:diggs}(b) shows the maximum diggs attained by stories in
our dataset vs rank of the submitter (the lower the rank, the more
successful the user). Slightly more than half of the stories came
from 14 top-rated users (rank$<25$) and 48 stories came from 45
low-rated users. The mean ``interestingness'' of the stories
submitted by the top-rated users is $600$, almost half the average
``interestingness'' of the stories submitted by low-rated users. A
second observation is that top-rated users are responsible for
multiple front page stories. A look at the statistics about top
users provided by Digg shows that this is generally the case: of the
more than $15,000$ front page stories submitted by the top 1020
users, the top $3\%$ of the users are responsible for $35\%$ of the
stories.

\section{Social networks and social filtering} If top-ranked users do not
submit the most interesting stories, why are they so successful? We
believe that social filtering play a role in promoting stories to
the front page. As we explained above, Digg's interface allows users
to designate others as ``friends'' and easily keep track of friends'
activities: the stories they have submitted, commented on or dugg.
We believe that users use this feature to filter the tremendous
number of new submissions on Digg. We show this by analyzing two
sub-claims: (a) \emph{users digg stories their friends submit}, and
(b) \emph{users digg stories their friends digg}.

\begin{figure*}[tbh]
\begin{tabular}{cc}
  \includegraphics[height=2.0in]{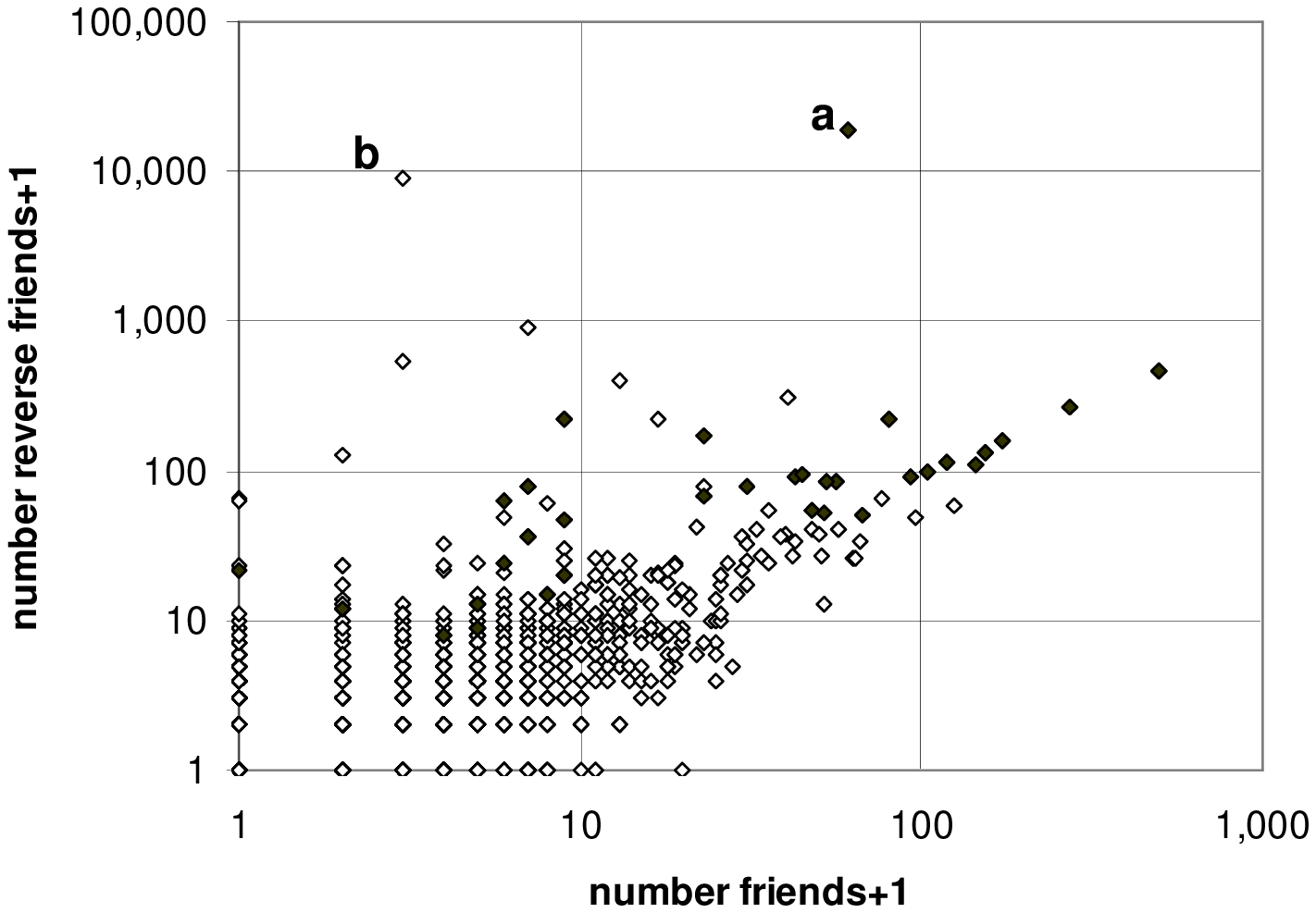}  &
  \includegraphics[height=2.0in]{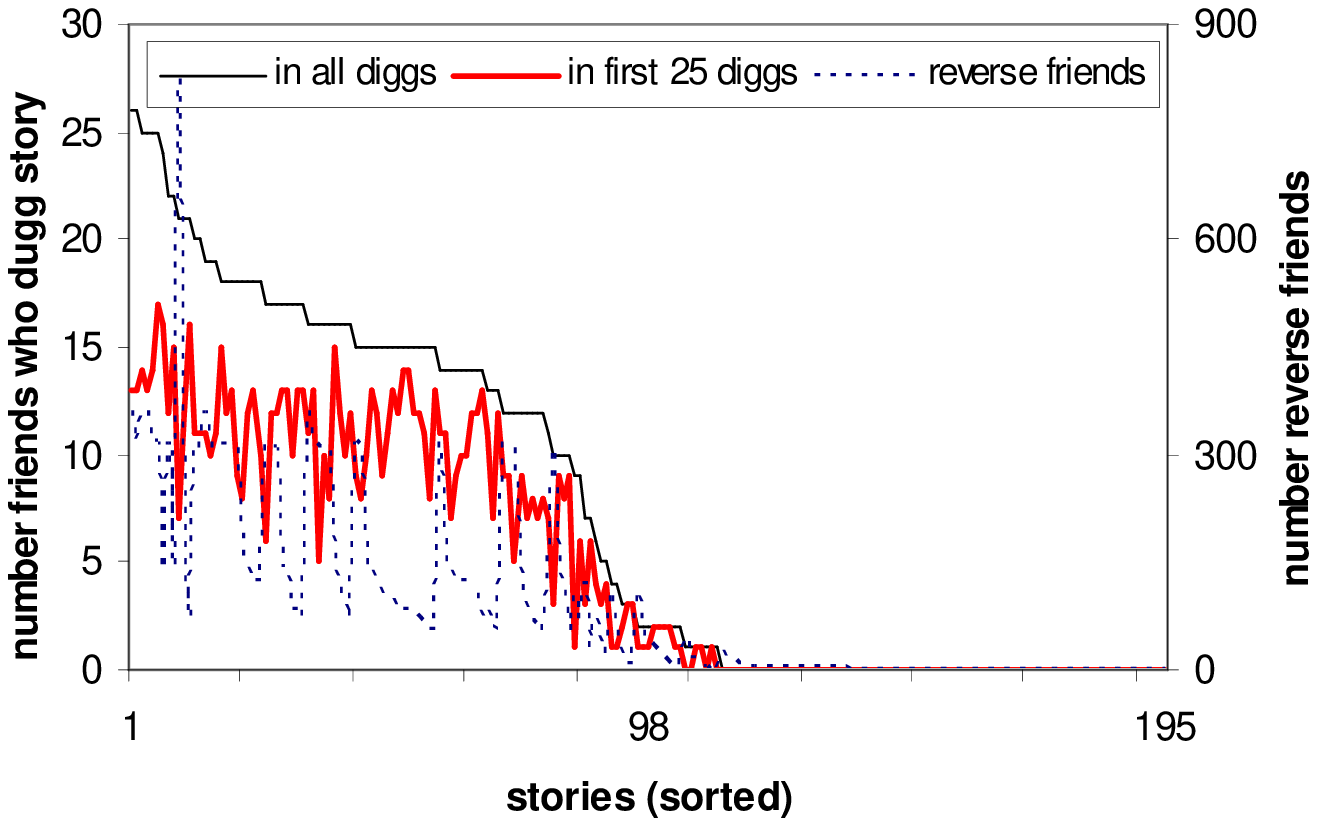}\\
  (a) & (b) \\
\end{tabular}
  \caption{(a) Scatter plot of the number of friends vs reverse friends for the top 1020 Digg users.
  (b) Number of diggers who are also among the reverse friends of the user who submitted the
story}\label{fig:scatterplot}
\end{figure*}

Note that the ``friend'' relationship is not symmetric: if user A
designates user B as a friend, user A can keep track of user B's
activities, but not vice versa. This makes A the \emph{reverse
friend} of B. \figref{fig:scatterplot}(a) shows the scatter plot of
the number of friends vs reverse friends of the top 1020 Digg users
as of May 2006. Black symbols correspond to the top 33 users. For
the most part, users appear to take advantage of Digg's social
networking feature, with the top users having bigger social
networks. Users below the diagonal are watching more people than are
watching them (fans), while users above the diagonal are being
watched by more other users than they are watching (celebrities).
Two of the biggest celebrities are users marked $a$ and $b$ on
\figref{fig:scatterplot}(a). These users are $kevinrose$ and
$diggnation$, respectively, one of the founders of Digg and a
podcast of the popular Digg stories.

\comment{
A user's success rate is defined as the fraction of the stories the
user submitted that have been promoted to the front page. We use the
statistics about the activities of the top 1020 users. In our
analysis, we only include users who have submitted 50 or more
stories (514 users). Users' mean success rate vs the size of their
social network is shown in \figref{fig:scatterplot}(b). Although the
error bars are large, there is a significant correlation between the
size of the user's social network (specifically,  number of reverse
friends) and user's success rate.

\begin{figure*}[tbh]
\begin{tabular}{cc}
  \includegraphics[width=3.0in]{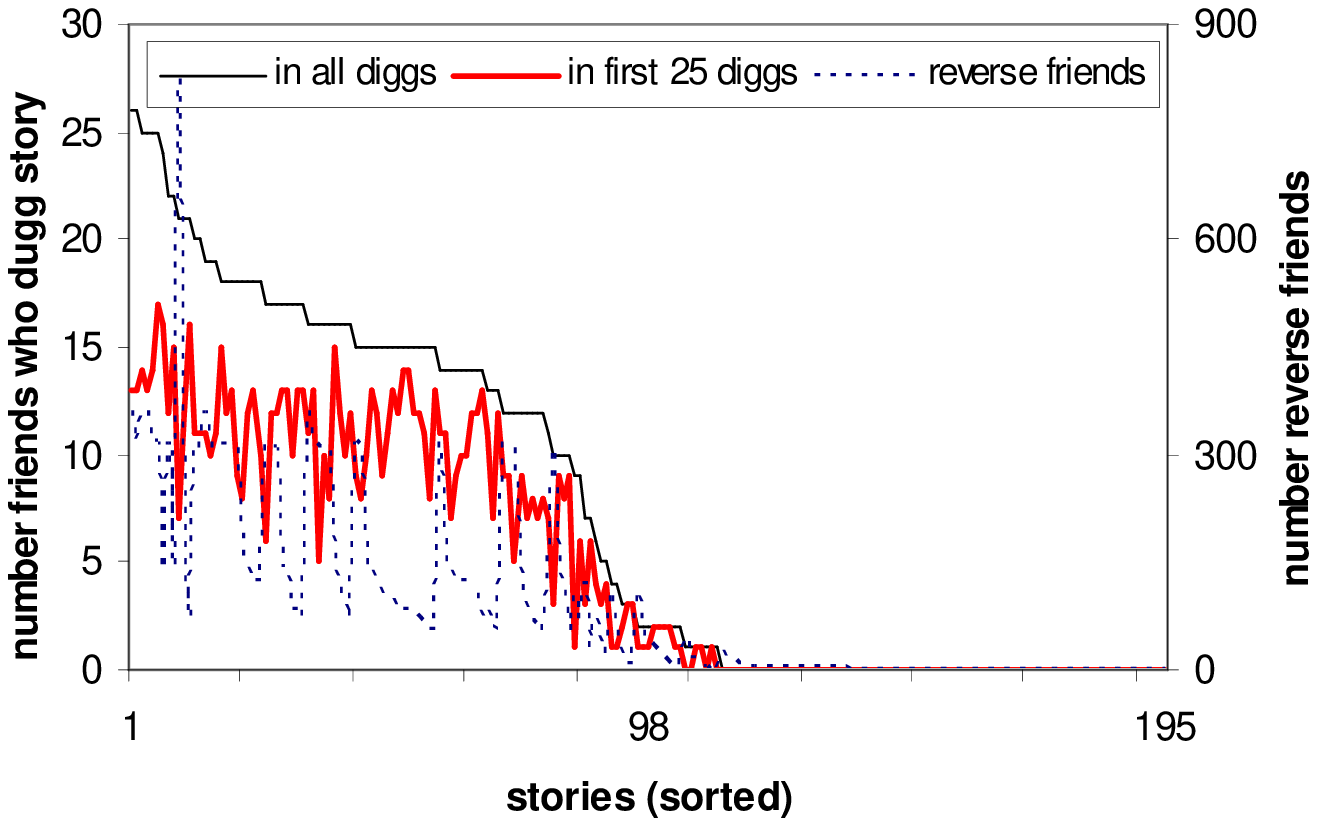}  &

\\
  (a) & (b) \\
\end{tabular}
  \caption{}\label{fig:diggs-history}

\end{figure*}

}

\subsection{Users digg stories their friends submit} In order to
show that users digg stories their friends submit, we used
 \emph{digg-with-history} wrapper to collect
195 front stories, each with a list of the first 216 users who dugg
the story ($15,742$ unique users total). The name of the submitter
is first on the list.

We can compare the list of users who dugg the story, or any portion
of this list, with the list of reverse friends of the submitter.
\figref{fig:scatterplot}(b) shows the number of diggers of a story
who are also among the reverse friends of the user who submitted the
story, for all 195 stories. Dashed line shows the size of the social
network (number of reverse friends) of the submitter. More than half
of the stories (102) were submitted by users with one or more
reverse friends, and the rest by unknown users.\footnote{These users
have rank $>1020$ and were not listed as friends of any of the 1020
users in our dataset. It is possible, though unlikely, that they
have reverse friends.} Thin solid line shows how many people who
list the submitter as a friend dugg the story within the first 215
diggs. All but two of the stories (submitted by SearchEngines with
21 reverse friends) were dugg by submitter's reverse friends. We use
simple combinatorics~\cite{Papoulis} to compute the probability that
$k$ of the submitter's friends could have dugg the story purely by
chance. The probability that after picking $n=215$ users randomly
from a pool of $N=15,742$ you end up with $k$ that came from a group
of size $K$ is $ P(k,n)={n\choose k} (p)^k (1-p)^{n-k}$, where
$p=K/N$. Using this formula, the probability (averaged over stories
dugg by at least one friend) that the observed numbers of friends
dugg the story by chance is $P=0.005$, making it highly
unlikely.\footnote{If we include in the average the two stories that
were not dugg by any of the submitter's friends, we end up with a
higher, but still significant P=0.023.} Moreover, users digg stories
submitted by their friends very quickly. The heavy solid line in
\figref{fig:scatterplot}(b) shows the number of reverse friends who
were among the first 25 diggers. The probability that these numbers
could have been observed by chance is even less --- $P=0.003$. We
conclude that users digg stories their friends submit. A consequence
of this conclusion is that users with active social networks are
more successful in getting their stories promoted to the front page.
We believe that this, coupled with the observation that top-ranked
users have larger social networks, explains their success.


\subsection{Users digg stories their friends digg} In the previous
section we showed that by enabling users to quickly digg stories
submitted by friends, social networks play an important role in
promoting content to the front page. Do social networks also help
users discover interesting stories that were submitted by unknown
users (users who are not listed as friends by anyone)? Top users are
very active. The top $3\%$ of the 1020 recently active Digg users in
our dataset is not only responsible for the disproportionate share
of front page stories, but they also submit more than $28\%$ of the
stories submitted by the group of 1020 users, and digg $11\%$ and
comment on $8\%$ of the stories dugg by and commented on by this
group. Once one of these well connected users diggs a story, others
within his or her social network will be more likely to read it
thanks to the user interface of Digg that quickly allows a user to
view stories dugg by friends.

\comment{
\begin{table}
  \centering
\begin{tabular}{|ll|c|c|c|c|c|c|}
  \hline
  & \textbf{diggers} & \textbf{m=1} & \textbf{m=6} & \textbf{m=16} & \textbf{m=26} & \textbf{m=36} & \textbf{m=46}
  \\ \hline
(a) & visible to friends & 26  & 60  & 88 & 96  & 100  & 101  \\
 (b) & dugg by friends & 9 & 18 & 29 & 35 & 39 & 47 \\
  (c) & probability & 0.002 & 0.017 & 0.047 & 0.069 & 0.061 & 0.071 \\ \hline
\end{tabular}
}

\begin{figure*}
\begin{tabular}{cc}
  \includegraphics[height=2.0in]{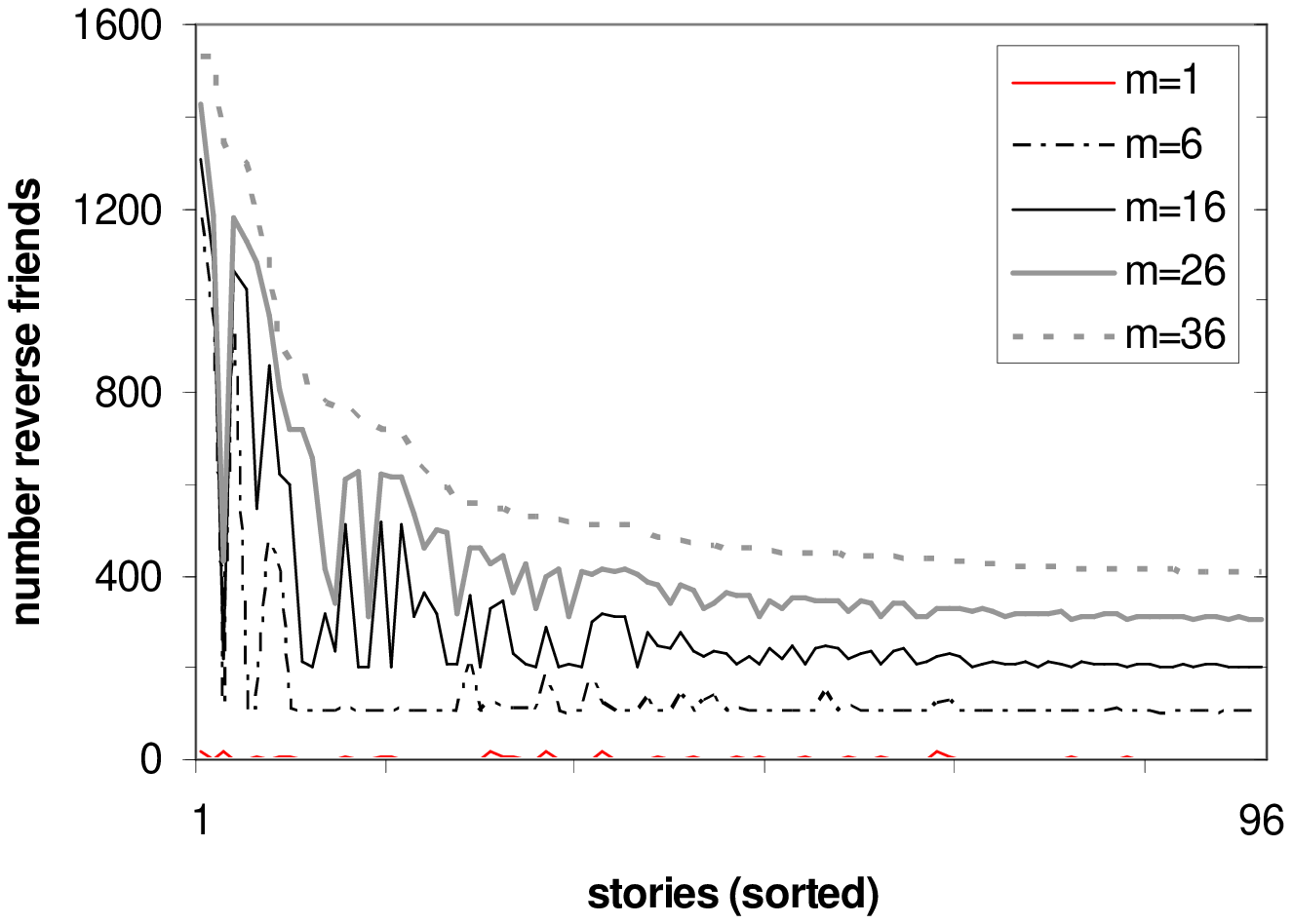}  &
  \includegraphics[height=2.0in]{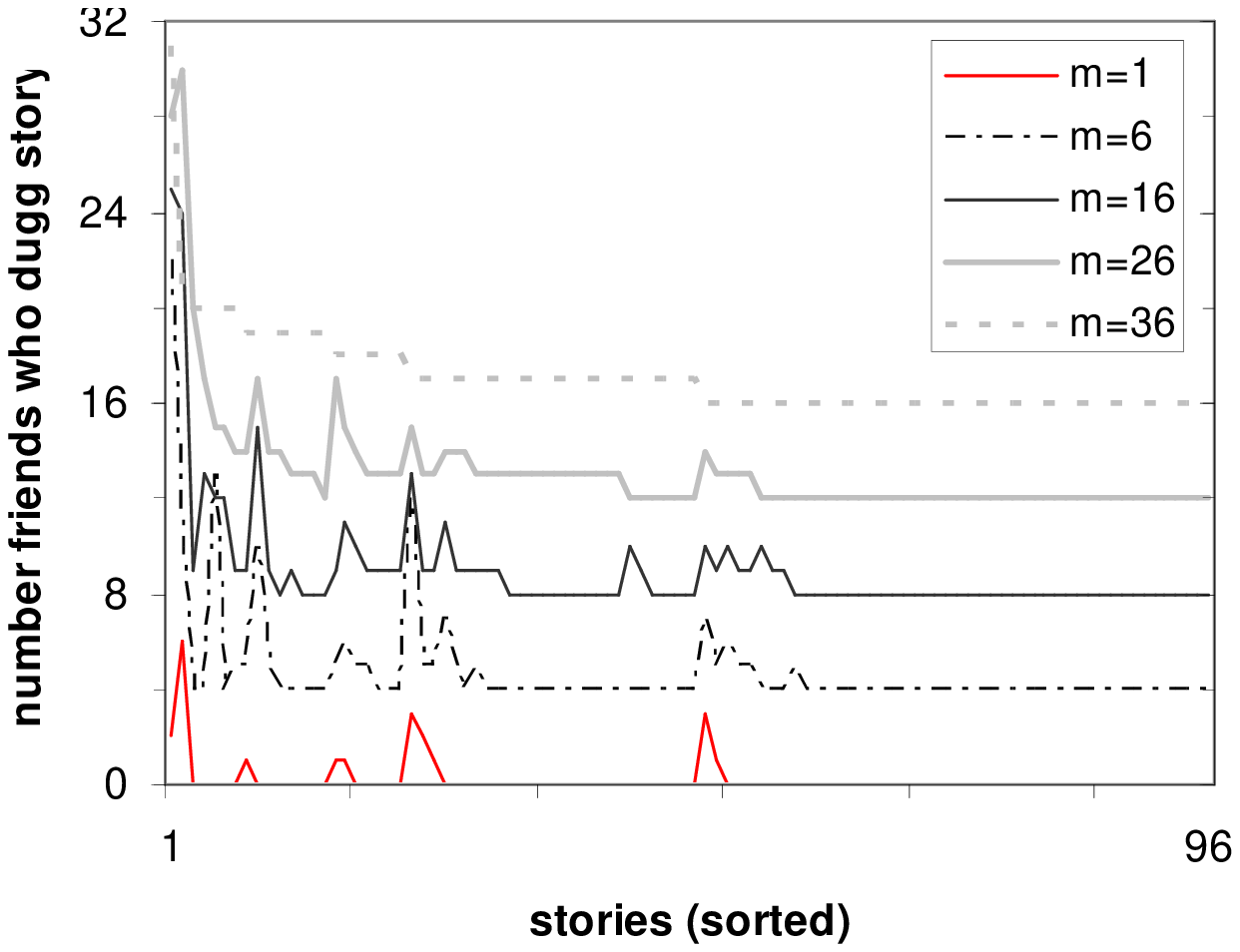}\\
  (a) & (b) \\
\end{tabular}
  \caption{(a) Number of reverse friends of the first $m$ diggers for the stories submitted by unknown users.
  (b) Number of friends of the first $m$ diggers who dugg the stories.}\label{fig:history-unknown}
\end{figure*}

\figref{fig:history-unknown} shows how digging activities of
well-connected users affect stories submitted by ``unknown'' users.
$m=1$ corresponds to the user who submitted the story, while $m=6$
corresponds to the story's submitter and the first five users to
digg it. Each line is shifted upward with respect to the preceding
line to aid visualization. Social networks increase story's
visibility. While at the time of submission, only 26 of the 101
stories were visible to other users within the submitter's social
network ($m=1$), by the time 25 others have dugg the story ($m=26$),
all the stories were visible to others through the friends
interface.

\begin{table*}
  \centering
\begin{tabular}{|ll|c|c|c|c|c|c|}
  \hline
  & \textbf{diggers} & \textbf{m=1} & \textbf{m=6} & \textbf{m=16} & \textbf{m=26} & \textbf{m=36} & \textbf{m=46}
  \\ \hline
(a) & visible to friends & 34 & 75  & 94 & 96  & 96  & 96  \\
 (b) & dugg by friends & 10 & 23 & 37 & 46 & 49 & 55 \\
  (c) & probability & 0.005 & 0.028 & 0.060 & 0.077 & 0.090 & 0.094 \\ \hline
\end{tabular}

  \caption{Number of stories posted by ``unknown'' users that were (a) made visible to other users through the
  digging activities of well-connected users, (b) dugg by friends of the first $m$ diggers within the next 25 diggs,
  and for the stories that were dugg by friends, (c) the average probability that the observed numbers of friends
  could have dugg the story by chance }\label{tbl:history-unknown}
\end{table*}

Do users digg stories dugg by friends? To answer this question we
look at the 25 diggs that come after the first $m$ diggs and see how
many of them come from friends of the $m$ diggers.  Only ten of the
stories were dugg by submitter's reverse friends. After five more
users dugg the stories ($m=6$), 75 became visible to others through
the friends interface, and of these 23 were dugg by friends. After
25 users have dugg the story, all 96 stories were visible through
the friends interface, and almost half of these were dugg by
friends. \tabref{tbl:history-unknown} summarizes the observations
and presents the probability that the observed numbers of friends
dugg the story by chance. The probabilities for $m=26$--$m=46$ are
above the $0.05$ significance level, and possibly reflect the
increased visibility the story receives once it makes it to the
front page. Although the effect is not quite as dramatic as one in
the previous section, we believe that the data shows that users do
use the friends interface to find new interesting stories.

\section{Comparison with Reddit}
Reddit~\cite{redditurl} is another social news aggregator that
allows users to submit and vote on stories. Stories that get enough
positive votes are then promoted to the ``hot'' page, Reddit's
version of the front page. Unlike Digg, Reddit does not have an
explicit social networking component which allows a user to track
friends's activities or browse another user's network of
friends.\footnote{Reddit added \emph{friends} feature in summer of
2006, a month after we collected data from the site. At the time of
the paper this feature is fairly rudimentary --- it simply allows
the user to quickly spot stories submitted by friends by
highlighting them.} Instead, Reddit lets users discover new
interesting stories through its recommendation system that uses
collaborative filtering to suggest stories that were liked by other
users with similar voting patterns. Alternately, a user can browse
through the newly submitted stories.

\begin{figure*}[tbh]
\begin{tabular}{cc}
  \includegraphics[width=3.0in]{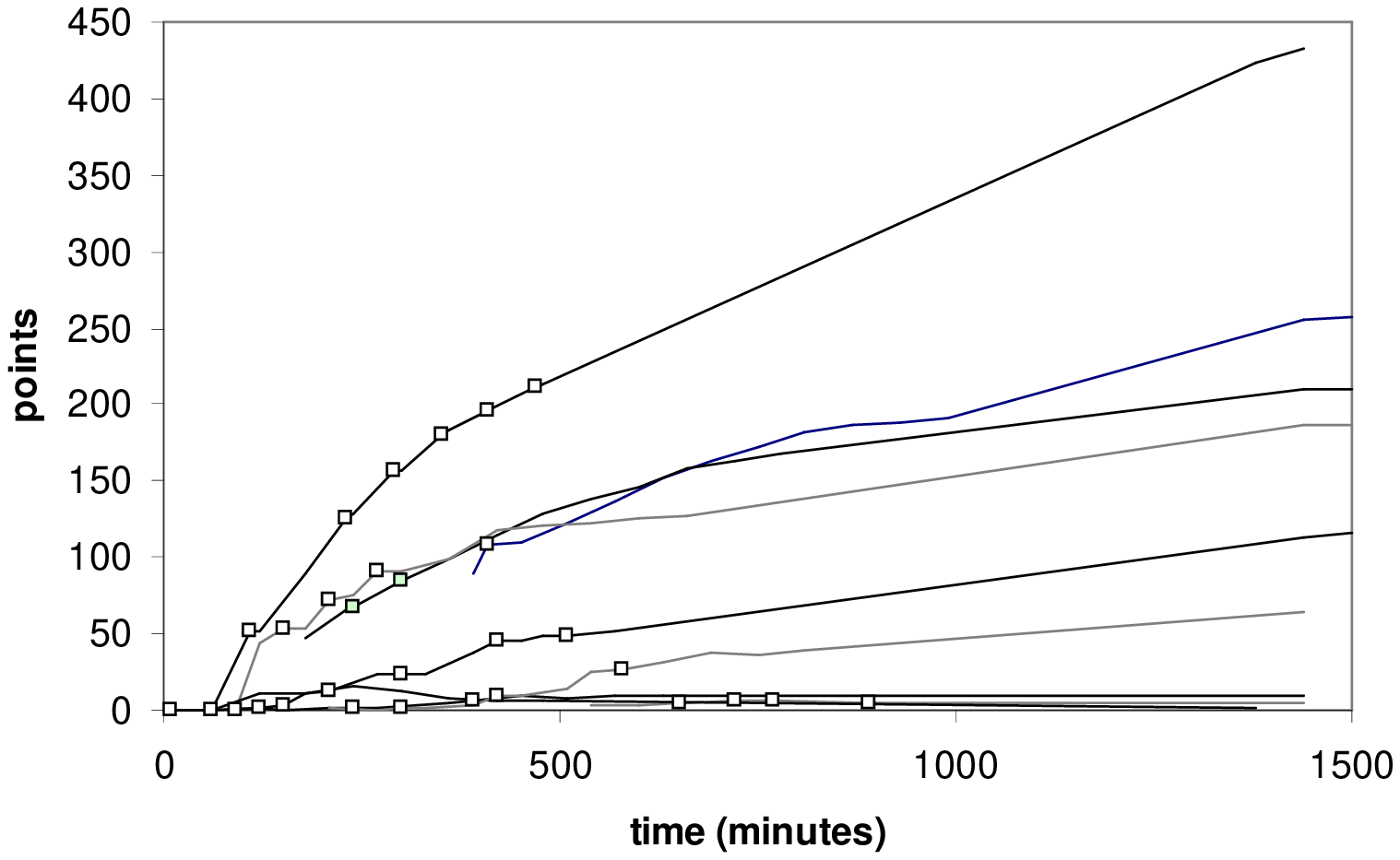}  &
\includegraphics[width=3.0in]{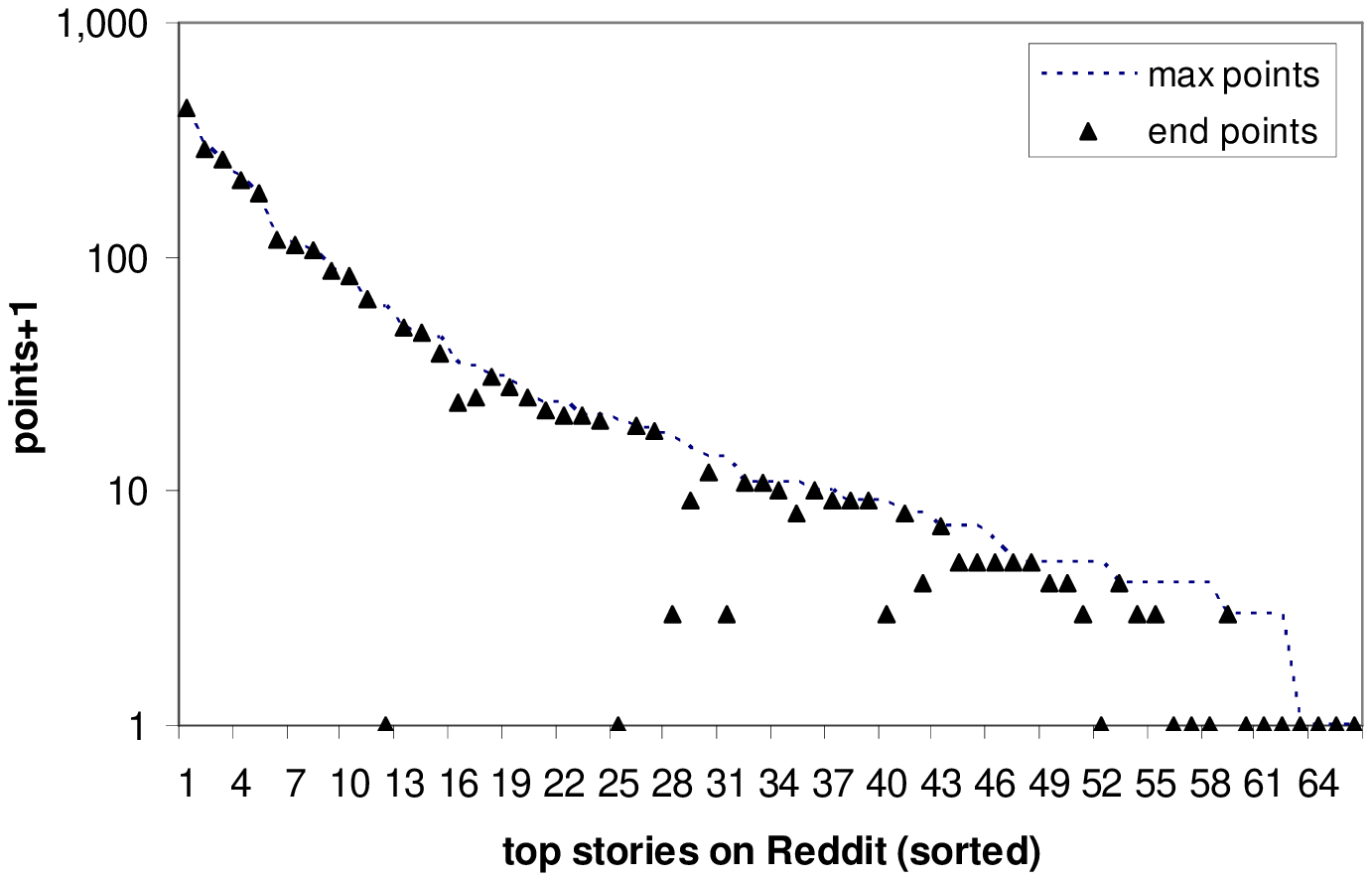} \\
  (a) & (b) \\
\end{tabular}
  \caption{(a) Points accumulated by stories on Reddit's hot page over a period of a day. Square markers show
  when the story also appeared on the \emph{new} page. (b) Maximum rating (in points) attained by Reddit stories over
  the tracking period compared to the rating they had at the end of the period. }\label{fig:reddit}
\end{figure*}

Our dataset consists of statistics extracted from Reddit's new and
hot pages over a period of two days in May, 2006. We identified 571
stories submitted by 350 users over the course of approximately a
day. Of these, 260 stories by 192 users also appeared on the hot
(front) page. \figref{fig:reddit}(a) shows how the number of points
accumulated by stories on Reddit's hot page changes with time. Note
that we were only able to track the stories for up to one day past
submission time. At first glance, dynamics looks similar to Digg.
Unlike Digg, however, a story often appears on the hot page at the
same time it appears on the new page (squares in
\figref{fig:reddit}). Also, unlike Digg, Reddit allows people to
vote stories down. \figref{fig:reddit}(b) shows the maximum number
of points achieved by Reddit stories over a period of about a day
and the points these stories had at the end of the period. One can
see that the ratings of a substantial number of stories dropped, in
many cases to zero, while other stories appeared on the front page
with very few points and never went anywhere.

We were unable to obtain data to measure the effectiveness of
Reddit's recommendation algorithm. We can only state that the
algorithm Reddit uses to promote stories (which must consider
actions of users reading and voting on recommended stories) appears
to be less effective than Digg's in that it allows many more
``uninteresting'' stories (whose ratings do not increase) to the
front page. This may account for the perception of Reddit as a
timelier source of news. On Digg, a story has to accumulate enough
votes before it is promoted, which takes time, while on Reddit, many
stories appear to be promoted soon after posting, regardless of how
many points they have accumulated. Although Reddit does not use the
friends system, thus eliminating the possibility of ``bloc voting,''
some users appear to be more successful than others in getting their
stories promoted. In our dataset, there were an average of 1.4 front
page stories per user on Reddit,  compared to 1.6 on Digg.

\section{Tyranny of the minority?}

The new social media sites offer a glimpse into the future of the
Web, where, rather than passively consuming information, users will
actively participate in creating, evaluating, and disseminating
information. Several such sites, Digg and Flickr, for example, allow
users to designate select users as ``friends'' and provide easy
interface to track friends' activities. Just as Google
revolutionized Web search by exploiting the link structure of the
Web --- created independently through the activities of many Web
page authors --- to evaluate the contents of information on Web
pages, social media sites show that it is possible to personalize
search through \emph{social filtering} that exploits the activities
of others in the user's social network.

We studied the role social networks and social filtering play in the
collaborative ranking of information. Specifically, we looked at how
news stories submitted to Digg are promoted to its front page.
Digg's goal is to have only the best of the stories featured on its
front page, and it employs aggregated opinion of thousands of its
users, rather than a few dedicated editors, to select the best
stories. Digg also allows users to create social networks by
designating others as friends and provides a seamless interface to
track friends' activities: what stories users in their social
network submitted, liked, commented on, etc. By tracking stories
over time, we showed that social networks play an important role in
collaborative information filtering.  Specifically, we showed (a)
users tend to like stories submitted by friends and (b) users tend
to like stories their friends read and like. This, in a nutshell, is
social filtering. Since some users are more active than others,
direct implementation of social filtering may lead to ``tyranny of
the minority,'' where a lion's share of front page stories come from
users with the most active social networks. This appears to be the
case for Digg, where visualization of the graph of mutual friends
shows a single cluster of composed of users among the 30 top-ranked
individuals, giving them an edge in future success. However,
precisely because these users are the most active ones, they play an
important role in filtering information and bringing to other
users's attention stories that would otherwise be buried in the
onslaught of new submissions.

\begin{figure}[tbh]
\includegraphics[width=3.0in]{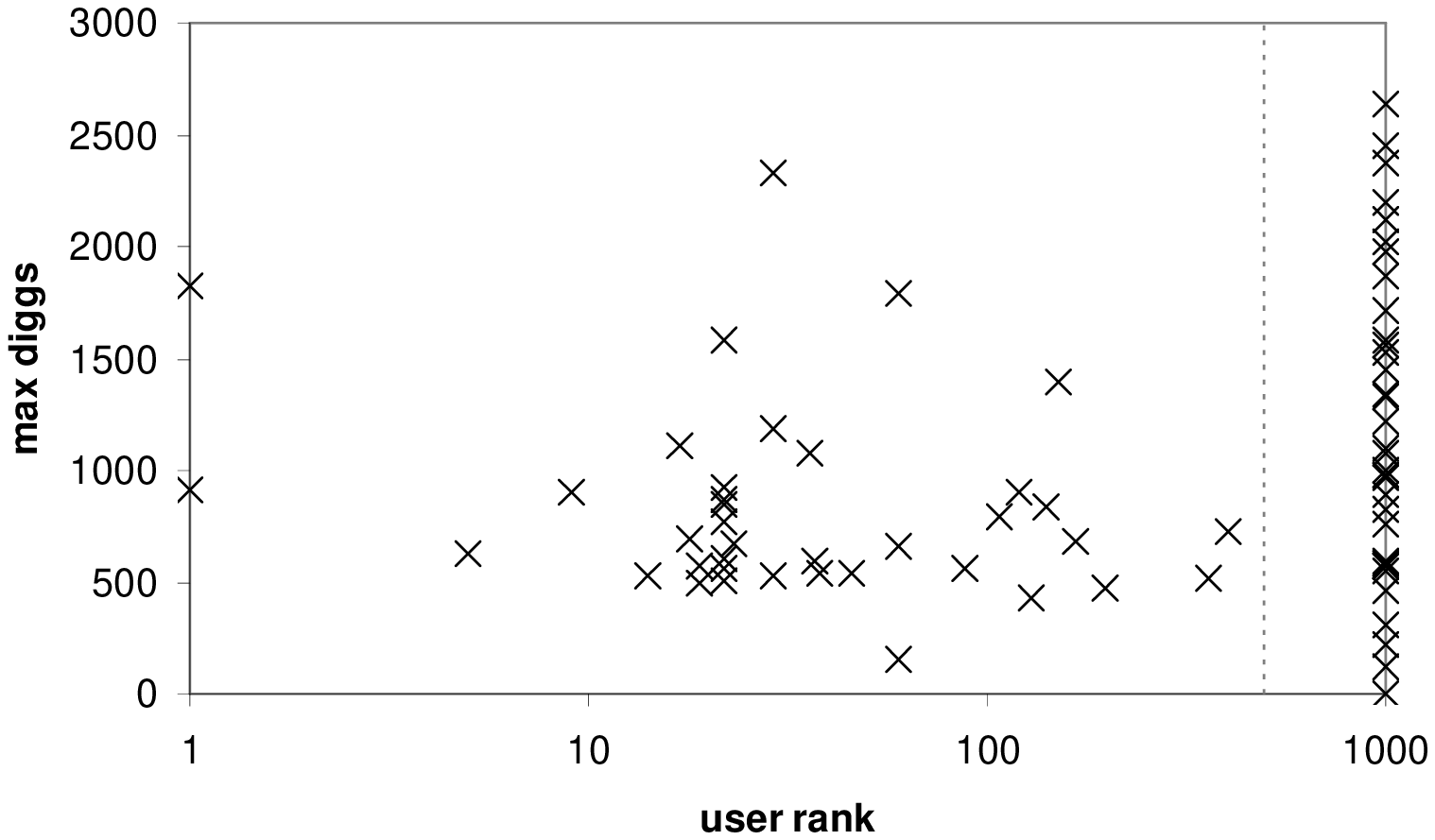}
  \caption{Maximum number of diggs attained by a front page story
  vs submitter's rank. Data was collected from stories submitted to Digg in early November 2006,
  after the change in the promotion algorithm. The vertical line divides the set in half.
  Symbols on the right hand axis correspond to low-rated users with rank$>1020$.
  }\label{fig:maxdiggs_new}
\end{figure}

Recently, a similar finding~\cite{taylorhaywardblog} resulted in a
controversy on Digg~\cite{USAToday}, in which users accused a
``cabal'' of top users of automatically digging each other's stories
in order to promote them to the front page. The resulting uproar
prompted Digg to change the algorithm it uses to promote stories. In
order to discourage what was seen as ``gaming'' the system through
``bloc voting,'' the new algorithm ``will look at the unique digging
diversity of the individuals digging the story''~\cite{diggblog}.
Preliminary results of the stories submitted in early November 2006
indicate that algorithm change did achieve the desired effect of
reducing the top user dominance on the front page. Our analysis of
the November data shows that of the 3015 stories submitted by 1866
users over about one day, 77 stories by 63 users were promoted to
the front page. \figref{fig:maxdiggs_new} shows the maximum number
of diggs received by these stories over a period of six days vs the
rank of the submitting user. Compared to \figref{fig:diggs}, front
page now has a greater diversity of users, with fewer users
responsible for multiple front page stories. In fact, in our data
set, there are 1.2 stories per submitting user, compared to 1.6
before. Although this may be seen as a positive development, the
change in the story promotion algorithm may have some unintended
consequences: it may, for example, discourage users from joining
social networks because their votes will be discounted. It is too
early to see what long term consequences, intended or not, the new
algorithm will have.

Rather than being a liability, however, social networks can be used
to personalize and tailor information to individual users, and drive
the development of new \emph{social search algorithms}. As Digg
matures, we expect different sub-communities to arise, each
representing users interested in a particular topic or a combination
of topics. A single user could belong to several different
communities, and use his or her social networks to find and filter
interesting new information. For example, Digg can create
personalized front pages for every user that are based on his or her
friends' readings. This will finally free individuals from ``tyranny
of the majority'' which results from viewing a common global front
page or best seller list.

In order to be effective for personalizing information, the social
networks created by users have to reflect their tastes and
interests. Some users appear to accumulate contacts for the sake of
having contacts, or reciprocate every request to be added to the
contacts list. On Flickr, for example, we have observed some users
with over 10,000 contacts. Publicly displaying one's tastes raises
many privacy issues, which have yet to be addressed. Promising or
perilous, social media appears to be the future of the Web.


\paragraph{Acknowledgements} This research is based on work
supported in part by the National Science Foundation under Award
Nos. IIS-0535182 and IIS-0413321. We are grateful to Dipsy Kapoor
for helping with data analysis, and to Fetch Technologies for
providing wrapper building and execution tools.

\bibliographystyle{plain}
\bibliography{../social}

\end{document}